\documentclass[10pt]{article}
\usepackage{epsfig,graphicx}
\usepackage{times,ch2005}

\def\beq{\begin{equation}}
\def\eeq#1{\label{#1}\end{equation}}
\def\eeqn{\end{equation}}

\def\beqa{\begin{eqnarray}}
\def\eeqa#1{\label{#1}\end{eqnarray}}
\def\eeqan{\end{eqnarray}}

\def\Dslash{\not{\hbox{\kern-4pt $D$}}}
\def\dslash{\not{\hbox{\kern-2pt $\del$}}}

\newcommand{\tev}{\ensuremath{\mathrm{\,Te\kern -0.1em V}}\xspace}
\newcommand{\gev}{\ensuremath{\mathrm{\,Ge\kern -0.1em V}}\xspace}
\newcommand{\mev}{\ensuremath{\mathrm{\,Me\kern -0.1em V}}\xspace}
\newcommand{\kev}{\ensuremath{\mathrm{\,ke\kern -0.1em V}}\xspace}
\newcommand{\ev}{\ensuremath{\mathrm{\,e\kern -0.1em V}}\xspace}
\newcommand{\gevc}{\ensuremath{{\mathrm{\,Ge\kern -0.1em V\!/}c}}\xspace}
\newcommand{\mevc}{\ensuremath{{\mathrm{\,Me\kern -0.1em V\!/}c}}\xspace}
\newcommand{\gevcc}{\ensuremath{{\mathrm{\,Ge\kern -0.1em V\!/}c^2}}\xspace}
\newcommand{\mevcc}{\ensuremath{{\mathrm{\,Me\kern -0.1em V\!/}c^2}}\xspace}
\def\mus  {\ensuremath{\rm \,\mus}\xspace}

\def\mus        {\ensuremath{\,\mu{\rm s}}\xspace}    %% microsecond

\begin{document}

%+ \Chapter{}
%+ {Instruction for producing Cherenkov 2005 proceedings}
%+ {Firstname~Yourname}

\Title{Analysis methods for Atmospheric Cerenkov Telescopes}
\bigskip

%+ \addcontentsline{toc}{chapter}{{\it Firstname~Yourname}}
%+ \index{author}{Yourname, P.} 

%%%%%%%%%%%%%%%%%%%%%%%%%%%%%%%%%%%%%
% Label to flag the first page of your contribution
% Replace Yourname by your name starting with a capital letter
%
\label{DenauroisStart}

%%%%%%%%%%%%%%%%%%%%%%%%%%%%%%%%%%%%%
% Your name
%
\author{ Mathieu de Naurois\index{de Naurois, M.} }

%%%%%%%%%%%%%%%%%%%%%%%%%%%%%%%%%%%%%
% Your address
%
\address{LPNHE\\
IN2P3 - CNRS - Universit\'es Paris VI/VIII \\
4 Place Jussieu
F-75252 Paris Cedex 05, France \\
}

\makeauthor\abstracts{Three different analysis techniques for Atmospheric Imaging System
are presented. The classical {\it Hillas parameters} based technique is shown to be
robust and efficient, but more elaborate techniques can improve the sensitivity
of the analysis. A comparison of the different analysis techniques shows that they use 
different information for gamma-hadron separation, and that it is possible
to combine their qualities.}

%%%%%%%%%%%%%%%%%%%%%%%%%%%%%%%%%%%%%%%%%%%%%%%%%%%%%%%%%%%%%%%%%%%%%%%%%
\section{Introduction}

From the beginning of ground based gamma ray astronomy, data analysis 
techniques were mostly based on the ``Hillas parametrisation'' \cite{Denaurois-Hillas-1985}
of the shower  images, relying on the fact that the gamma-ray images in the camera focal plane are, to 
a good approximation, elliptical in shape. More elaborate analysis techniques were 
pioneered by the work of the CAT collaboration on a model analysis technique, where the shower images 
are compared to a more realistic pre-calculated model of image. Other analysis techniques, such
as the {\it 3D Model analysis} were developed more recently with the start of the third-generation
telescopes. The {\it 3D Model analysis} is, for instance, based on the assumption  of a 3 
dimensional elliptical shape of the photosphere.

These analysis techniques are complementary in many senses. We will show that they are sensitive
to different properties of the shower, and can therefore be used to cross-check the analysis 
results or be combined together to improve the sensitivity. No analysis is currently really
winning the race, and there is much space for further improvements.

%%%%%%%%%%%%%%%%%%%%%%%%%%%%%%%%%%%%%%%%%%%%%%%%%%%%%%%%%%%%%%%%%%%%%%%%%
\section{Hillas-parameter based analysis}

\subsection{Introduction}

In a famous paper of 1985\cite{Denaurois-Hillas-1985}, M. Hillas proposed to reduce
the image properties to a few numbers, reflecting the modelling of the image
by a two-dimensional ellipse. These parameters, shown on figure \ref{fig:Denaurois-figHillas},
are usually:

\begin{figure}[htb]
\begin{minipage}[b]{0.54\linewidth}
\begin{itemize}
\item length $L$ and width $w$ of the ellipse
\item size (total image amplitude)
\item nominal distance $d$ (angular distance between the centre of the camera
and the image centre of gravity)
\item azimuthal angle of the image main axis $\phi$
\item orientation angle $\alpha$
\end{itemize}
\end{minipage}
\hspace{0.1cm} % To get a little bit of space between the figures
\begin{minipage}[b]{0.45\linewidth}
\begin{center}
\epsfig{file=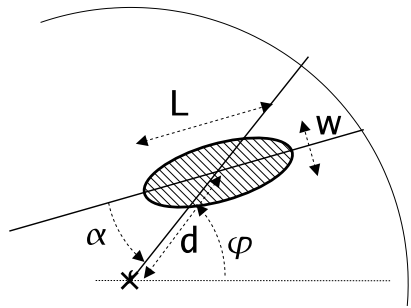,height=4cm}
\caption{Geometrical definition of the Hillas Parameters.}
\label{fig:Denaurois-figHillas}
\end{center}
\end{minipage}
\end{figure}

\subsection{Single telescope reconstruction}

In single telescope observations, the shower direction was estimated from the Hillas
parameters themselves (and in particular from the image {\it length} and {\it size}), 
either with lookup tables or with ad-hoc analytical functions. But the choice of
a symmetrical parametrisation of the shower led to degenerate solutions, on each
side of the image centre of gravity along the main axis. 

In order to break this degeneracy, other parameters --- based in particular
on the third order moments --- were added later.

The shower energy is usually estimated with a similar technique, from the image
{\it size} and {\it nominal distance}.

\subsection{Stereoscopic reconstruction}

The stereoscopic imaging technique, pioneered by HEGRA\cite{Denaurois-Daum-1997}, provides a
simple geometric reconstruction of the shower: the source direction is given by the intersection 
of the shower image main axes in the camera, and the shower impact is obtained
in a similar manner. The energy is then estimated from a weighted average of each single
telescope energy reconstruction.

\subsection{Gamma-hadron separation}

The Hillas parameters not only allow to reconstruct the shower parameters, but also can provide
some discrimination between $\gamma$ candidates and the much more numerous hadrons.
Several technique were developed, exploiting to an increased extent the existing correlation
between the different parameters (e.g. {\it Supercuts}\cite{Denaurois-Reynolds-1993}, 
{\it Scaled Cuts}\cite{Denaurois-Daum-1997} and {\it Extended Supercuts}\cite{Denaurois-Mohanty-1998}).
We will use here the {\it Scaled Cuts} technique, in which the actual image width ($w$) and length ($l$) 
are compared to the expectation value and variance obtained from simulation as a function of the 
image charge $q$ and reconstructed impact distance $\rho$, expressed by two normalised parameters
{\it Scaled Width}(SW) and {\it Scaled Length}(SL):

\begin{equation}
SW = \frac{w(q,\rho) - \left<w(q,\rho)\right>}{\sigma_w(q,\rho)},\quad SL = \frac{l(q,\rho) - \left<l(q,\rho)\right>}{\sigma_l(q,\rho)}
\end{equation}
 
These parameters have the noticeable advantage of being easily combined in stereoscopic observations
in {\it Mean Scaled Width} and {\it Mean Scaled Length}:

\begin{equation}
MSW = \frac{\displaystyle \sum_{tels} SW}{\sqrt{ntels}},\quad MSL = \frac{\displaystyle \sum_{tels} SL}{\sqrt{ntels}}
\end{equation}

From simulations, one can show that the {\it Mean Scaled Width} and {\it Mean Scaled Length}
are almost uncorrelated for $\gamma$ candidates ($\rho = 0.15 \pm 0.01$) and can therefore be 
combined in a single variable {\it Mean Scaled Sum} (MSS) : $MSS = (MSW + MSL)/\sqrt 2$.

% extended supercut = Hillas parameters rescaled by total charge

%%%%%%%%%%%%%%%%%%%%%%%%%%%%%%%%%%%%%%%%%%%%%%%%%%%%%%%%%%%%%%%%%%%%%%%%%
\section{Model analysis}

\subsection{Introduction}

The {\it Model Analysis}, introduced by the CAT collaboration\cite{Denaurois-Piron-2001} (with a single
telescope) and  further developed in the H.E.S.S. collaboration\cite{Denaurois-DeNaurois-2003}, is based
on the pixel-per-pixel comparison of the shower image with a template generated by
a semi-analytical shower development model. 
The event reconstruction is based on a maximum likelihood method which uses all available pixels in the camera,
without the requirement for an image cleaning.
The probability density function of observing a signal $S$ in a given pixel, given an expected amplitude $\mu$, 
a fluctuation of the pedestal $\sigma_p$ (due to night sky background and electronics) 
and a fluctuation of the single photoelectron signal (p.e.) $\sigma_s\approx 0.4$ (PMT resolution) is given by
the formula:

\begin{equation}
P(S|\mu,\sigma_p,\sigma_s) = \sum_{n=0}^\infty \frac{e^{-\mu} \mu^n}{n!\sqrt{2\pi(\sigma_p^2 + n \sigma_s^2)}} \exp \left(
- \frac{(S - n)^2}{2(\sigma_p^2 + n \sigma_s^2)} \right) 
%&\approx& \frac{1}{\sqrt{2\pi\left[\mu(1+\sigma_s^2) + \sigma_p^2 \right]}} \exp \left[
%- \frac{(S - \mu)^2}{2\left(\mu(1+\sigma_s^2) + \sigma_p^2 \right)} \right]
\end{equation}

\noindent The log-likelihood function
$ {\cal L} = 2 \sum_{\mathrm{pixel}} \log \left[P_i(S_i|\mu,\sigma_p,\sigma_s)\right]$
is then maximised to obtain the primary energy, direction and impact. In contrast to
the Hillas Analysis technique, the shower reconstruction works
in an identical way for a single telescope or for a stereoscopic array.

In stereoscopic observations, the {\it Model analysis} uses by its nature the correlations 
between the different images to find the best source direction and position, but in contrast
to the Hillas analysis, it doesn't take into account the shower fluctuations.

\subsection{Gamma-hadron separation}

In the {\it Model analysis}, the separation between the $\gamma$ candidates and the hadrons
is done by a {\it goodness-of-fit} ($G$) variable. The average value of the log-likelihood can
be calculated analytically:

\begin{eqnarray}
\left<\ln {\cal L}\right> &=& \sum_{pixel\,i} \int_{S_i} P(S_i|\mu_i,\sigma_{p_i},\sigma_{s_i}) \times \ln P(S_i|\mu_i,\sigma_{p_i},\sigma_{s_i}) \, dS_i \nonumber \\
&=& \sum_{pixel\,i}\left[1 + \ln(2\pi) + \ln\left( \sigma_{p_i}^2  + \mu_i\times (1 + \sigma_{s_i}^2)\right)\right]  
\end{eqnarray}

The variance of $\ln {\cal L}$ being close to 2, we define the {\it goodness-of-fit} $G$ as a normal variable ($N_\mathrm{dof}$ is the number of degrees of freedom):

\begin{equation}
G = \frac{\left<\ln {\cal L}\right> -\ln {\cal L}}{\sqrt {2\times N_\mathrm{dof}}}
\end{equation}

%%%%%%%%%%%%%%%%%%%%%%%%%%%%%%%%%%%%%%%%%%%%%%%%%%%%%%%%%%%%%%%%%%%%%%%%%
\section{3D Model analysis}

\subsection{Introduction}

The third and most recent analysis presented here, the {\it 3D Model Analysis}\cite{Denaurois-Lemoine-ch2005},
is a kind of 3 dimensional generalisation of the Hillas parameters: the shower is modelled as
a Gaussian {\it photosphere} in the atmosphere (with anisotropic light angular distribution),
which is then used to predict --- with a line of sight path integral --- the collected light in each pixel.
A comparison of the actual image to the predicted one (with a log-likelihood function) allows eight shower 
parameters to be reconstructed: {\it mean altitude}, {\it impact and direction}, {\it 3D width and length} and {\it luminosity}.

\begin{figure}[htb]
\begin{minipage}[b]{0.54\linewidth}
Reconstructed parameters:
\begin{itemize}
\setlength{\itemindent}{-1em}
\item {\it mean altitude}
\item {\it impact}
\item {\it direction}
\item {\it 3D width and length}
\item {\it luminosity}
\end{itemize}
\end{minipage}
\hspace{0.1cm} % To get a little bit of space between the figures
\begin{minipage}[b]{0.45\linewidth}
\begin{center}
\epsfig{file=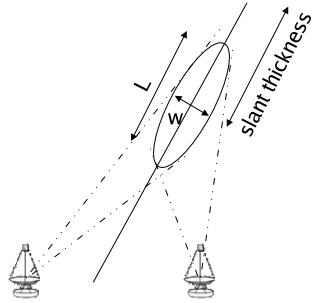,height=4cm}
\caption{Geometrical definition of the 3D Model Parameters.}
\label{fig:Denaurois-fig3DModel}
\end{center}
\end{minipage}
\end{figure}

\subsection{Gamma-Hadron separation}

The 3D-Model analysis relies on the strong assumption of a rotational symmetry,
which is used to reject about 70\% of the hadrons during the fit procedure.
For the remaining events, the most discriminating parameter between the $\gamma$ candidates and the hadrons is
found to be the shower width, as the hadronic showers are typically much wider 
than the electromagnetic ones. The shower width, expressed in units of radiation 
length, is found by simulation to be proportional to the slant thickness. This is 
is used to define a zenith angle independent {\it Reduced width} parameter:

\begin{equation}
\omega = \frac{w \times \rho{z_\mathrm{max}}}{\mathrm{thickness}}
\end{equation}

For simplicity, we will use here a {\it Rescaled width} ($Wr_{3D}$) parameter 
constructed from the {\it Reduced width} with a fixed offset and a
fixed ratio, to be a Gaussian distributed variable with mean 0 and RMS 1. 

%%%%%%%%%%%%%%%%%%%%%%%%%%%%%%%%%%%%%%%%%%%%%%%%%%%%%%%%%%%%%%%%%%%%%%%%%
\section{Comparison}

In this section we compare the properties of the three analyses presented above.
For that purpose, we will use two data sets:
\begin{itemize}
\item a real data set of 10 live hours obtained
by H.E.S.S. on the Crab Nebula in 2004 with 3 telescopes.
\item a simulation data set at zenith.
\end{itemize}

\subsection{Selection variables and efficiency}

\begin{figure}[htb]
\begin{center}
\epsfig{file=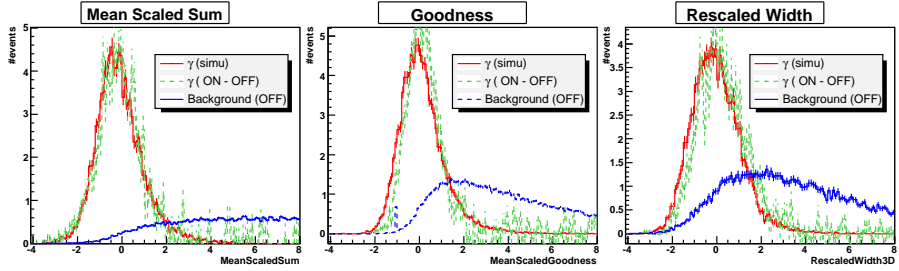,height=3.6cm}
\caption{Distribution of discriminating parameters obtained for each analysis for simulated $\gamma$'s, real OFF data
and real $\gamma$ obtained from ON-OFF data. Left: {\it Mean Scaled Sum}, Middle: Model {\it Goodness of Fit}, Right: 3D Model {\it Rescaled Width}.}
\label{fig:Denaurois-figParameters}
\end{center}
\end{figure}

The distributions of the three discriminating parameters ({\it Mean Scaled Sum}, 
{\it Goodness} and {\it Rescaled Width}) are shown in figure \ref{fig:Denaurois-figParameters} 
for simulated $\gamma$'s, real OFF data and real ON-OFF data. For the three analyses,
the real ON-OFF distributions are well reproduced by the simulation and are compatible
with normal variables. The OFF distributions have a different shape, indicating that
a cut $V\leq V_{\mathrm{max}}$ on these variables can be used.

\begin{figure}[htb]
\begin{center}
\epsfig{file=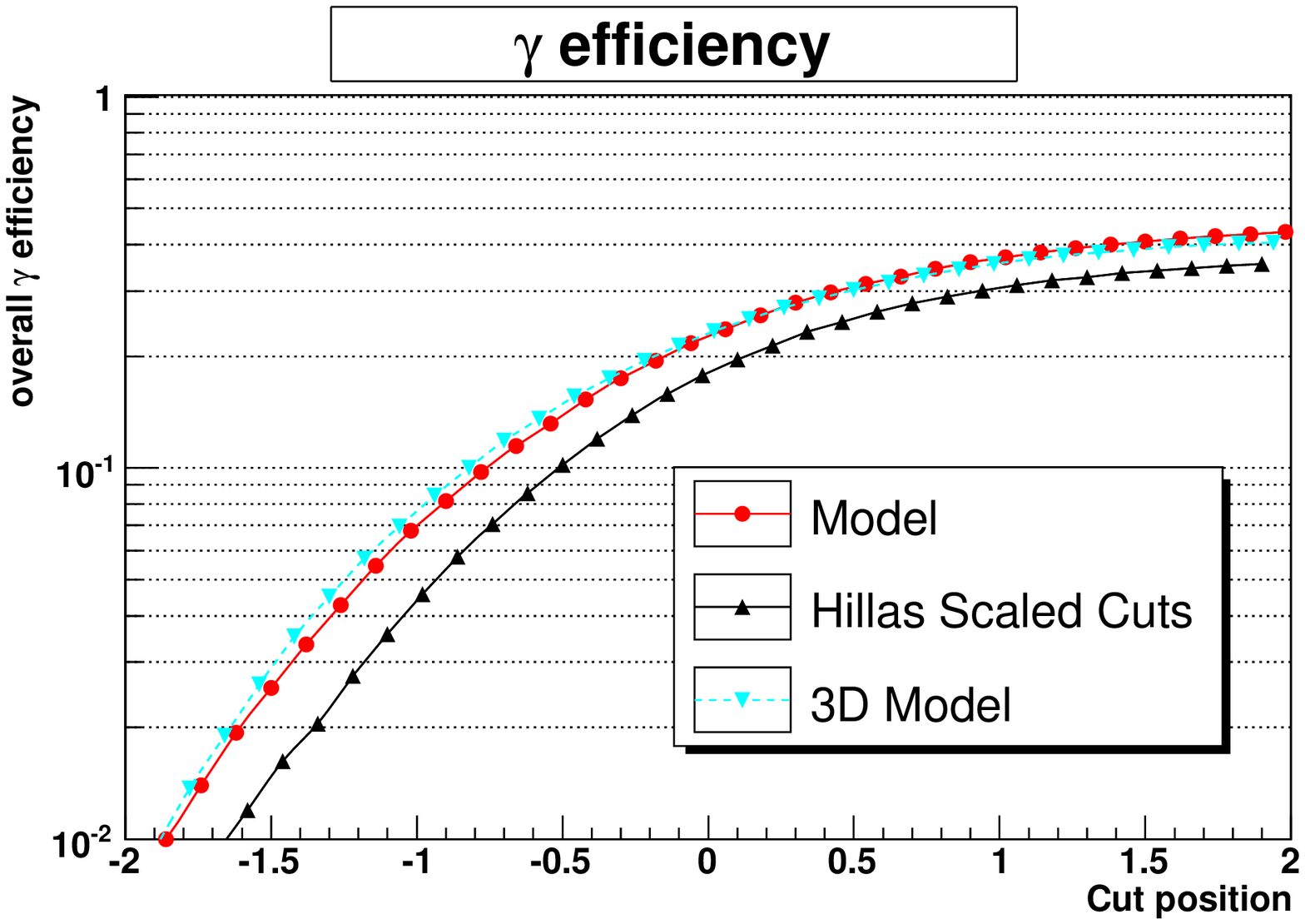,width=6cm}\hspace{-0.2cm}
\epsfig{file=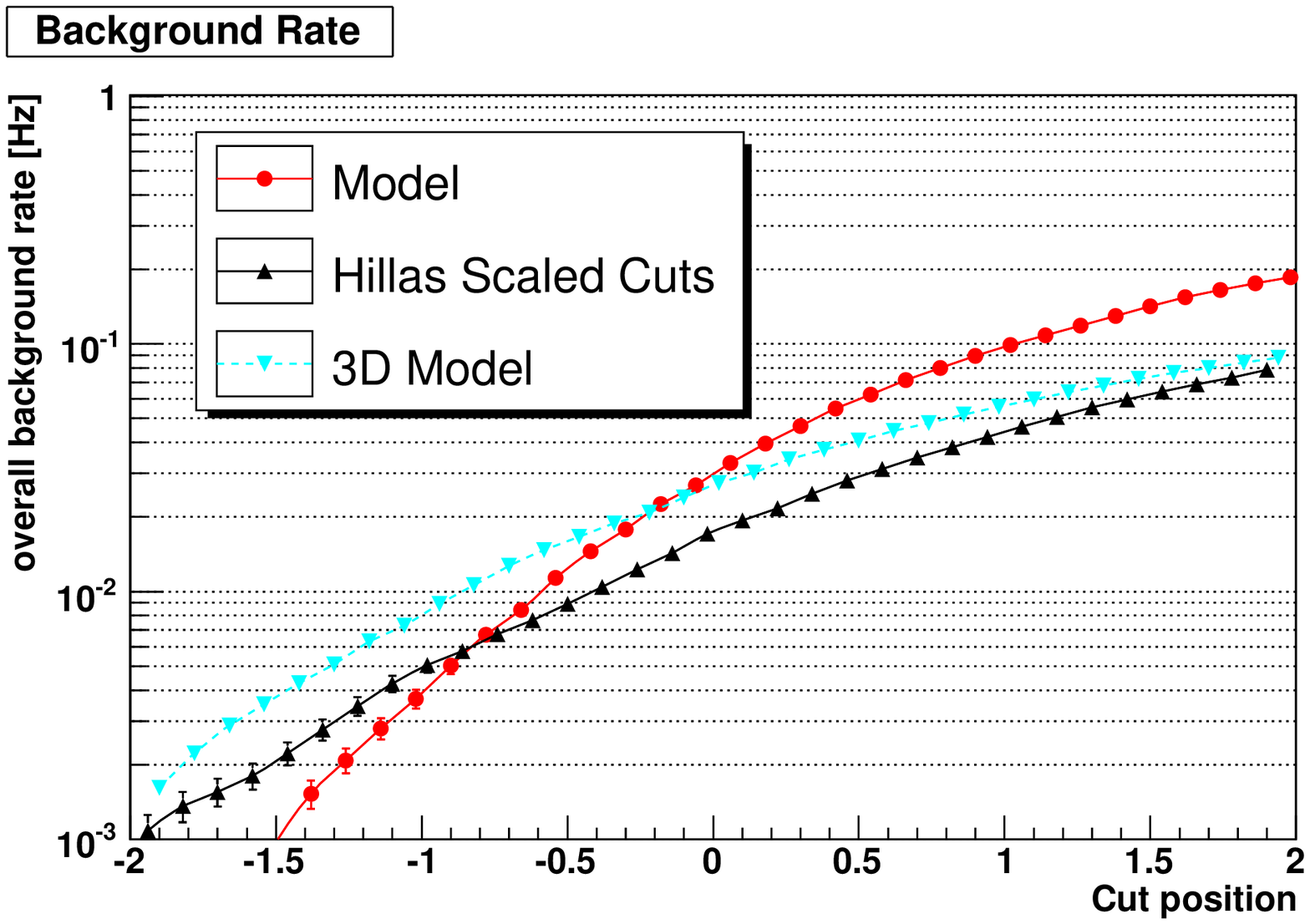,width=6cm}
\caption{Left: efficiency of the three analyses to simulated $\gamma$ with respect to the cut position. Right: remaining
background rate (from real data) for the same cuts.}
\label{fig:Denaurois-figEfficiency}
\end{center}
\end{figure}

Figure \ref{fig:Denaurois-figEfficiency} shows the efficiency of the three analyses for a point
source at zenith, including the reconstruction efficiency and the efficiency of a $\theta^2\leq 0.02$ cut,
as a function of the position of $V_{\mathrm{max}}$. The three curves on the left
panel are very similar in shape, reflecting the fact that the selection variables all have a normalised
Gaussian distribution. The Model and 3D Model analyses have a $\sim 20\%$ higher efficiency for $\gamma$-rays 
compared to the Hillas parameters based method, partially thanks to a higher reconstruction efficiency
and partially due to a slightly better angular resolution. They also keep more background events,
which in turns leads to very similar sensitivities (table \ref{tab:Denaurois-CombinedAnalysis}).

\subsection{Discriminating parameters correlations}

\begin{figure}[htb]
\begin{center}
\epsfig{file=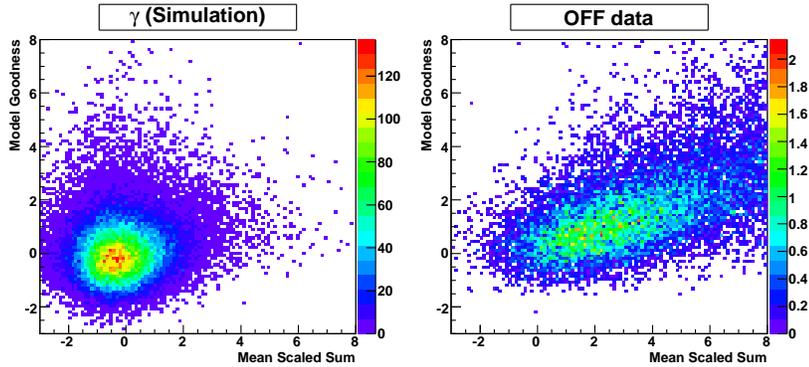,height=5cm}
\caption{Correlation of selection variables for two different analyses
on the same events.}
\label{fig:Denaurois-figCorrelation}
\end{center}
\end{figure}

Since the three analysis presented here have similar sensitivities and $\gamma$ efficiencies,
one would expect to see a strong correlation between the discriminating variables.
Figure \ref{fig:Denaurois-figCorrelation} shows the distribution of Model goodness-of-fit
versus Hillas Mean Scaled Goodness for simulated $\gamma$'s (left) and real OFF data (right).

The surprise is that there is almost no correlation between these variables
for simulated $\gamma$'s (correlation factor $\rho\approx 0.13$) whereas there is
some for the OFF data ($\rho\approx 0.53$). The same effect is seen when comparing
the Hillas analysis to the 3D Model, or the 3D Model with the Model. The obtained
correlation factors are summarised in the table \ref{tab:Denaurois-CorrelationTable}.

\begin{table}[htbp]
\begin{center}
\begin{tabular}{|c|c|c|}
\hline
Coefficient & $\gamma$ (Simulation) & OFF data \\ 
\hline
Model / Hillas & $0.126\pm 0.004$ & $0.53\pm 0.02$ \\
3D Model / Hillas & $0.03\pm 0.004$ & $0.51\pm 0.02$ \\
3D Model / Model & $0.120\pm 0.004$ & $0.35\pm 0.02$ \\
\hline
\end{tabular}
\caption{Correlation between the discriminating variables of the three 
describes analyses, for simulated $\gamma$'s and real OFF data.}
\label{tab:Denaurois-CorrelationTable}
\end{center}
\end{table}

The reason of this lack of correlation is not perfectly clear yet, but it must certainly
be due to the fact that the analyses are sensitive to different shower properties:

\begin{itemize}
\item The Hillas - {\it Scaled cuts} analysis takes into account the shower development
fluctuations (in the construction of the scaled cuts tables), but doesn't take into account
the details of the light distribution inside the shower image (and in particular its asymmetry),
nor the correlations between the images
\item The Model analysis takes into account the details of the light distribution and the correlations
between the images but not the shower fluctuations
\item The 3D Model analysis takes into account the correlations between the images, and some aspects of
the details of light distribution as well as the shower fluctuations (through their effect on the shower
length and width).
\end{itemize}

This also shows that none of these analyses completely exploits the available 
information, and that significant improvements can be achieved. In particular,
it's is possible to combine the selection variables by simpling adding them.
We define two new discriminating variables $V2$ and $V3$:

\begin{equation}
V2 = \frac{MSS + G}{\sqrt 2},\quad\quad V3 = \frac{MSS + G + Wr_{3D}}{\sqrt 3}
\end{equation}

The results obtained on the Crab data sample with the three analysis, using the
same cut position\footnote{Using the same cut position allows
an easy comparison of the different analyses. However, since they have different
rejection performances, the optimal cut position differs from one analysis to the other. 
The purpose here is not to state that one analysis is more efficient than the
other ones, but only to show that their respective performances can be combined.}
 ($V2\leq 0.8$ or $V3\leq 0.8$ ), a 60 photoelectrons cut and a 2 degrees Nominal 
distance cut (to remove the events close to the edge of the camera) are shown
together with the results of the {\it combined cuts} $V2$ and $V3$ in table
\ref{tab:Denaurois-CombinedAnalysis}. It should be noted that the average zenith angle
of the Crab dataset is roughly $40^\circ$ and that this dataset was taken with 3 telescope,
so the actual $\gamma$ efficiency and hadron rejection cannot be directly compared
to the values obtained on simulation in figure \ref{fig:Denaurois-figEfficiency}.

\begin{table}[htbp]
\begin{center}
\begin{tabular}{|c|c|c|c|c|c|c|}
\hline
Analysis & ON & OFF & $\#\gamma$ & $\sigma$ & S/B & $\sigma_{10}$ \\ 
\hline
Model ($G\leq 0.8$) & 2725 & 481 & 2244 & 41.6 & 4.5 & 6.5 \\
Hillas ($MSS\leq 0.8$)  & 1979 & 254 & 1725 & 38.9 & 7.0 & 6.6 \\
3D Model ($Wr_{3D}\leq 0.8$)  & 1908 & 309 & 1599 & 35.8 & 5.2 & 5.7 \\
\hline
Model/Hillas ($V2 \leq 0.8$) & 2587 & 225 & 2362 & 48.2 & 10.5 & 9.1 \\
3 analysis ($V3 \leq 0.8$) & 2197 & 165 & 2032 & 45.6 & 13.1 & 9.5 \\
\hline
\end{tabular}
\caption{Results of the three analyses and the combined analysis on a Crab data set. 
For each analysis, we quote the number of ON and OFF events, the number of $\gamma$, the significance, the signal
over background ratio and the significance obtained for a 10 times fainter source.}
\label{tab:Denaurois-CombinedAnalysis}
\end{center}
\end{table}

Even for a strong source, the significance is noticeably increased when combining the analyses
together. For faint sources, the gain in significance is almost $40\%$, due to a two times
better background rejection for roughly the same $\gamma$ acceptance.

\subsection{Resolution}

Comparing the angular or energy resolution between different analysis is always
a tricky business, since the values obtained depend in particular on the selection 
criteria which are specific to each analysis. 
The clean way to do it is to use a common sample of events. We use
here the events selected with $V3\leq 0.8$, which does not favour any analysis
with respect to the others.

\begin{figure}[htb]
\begin{center}
\epsfig{file=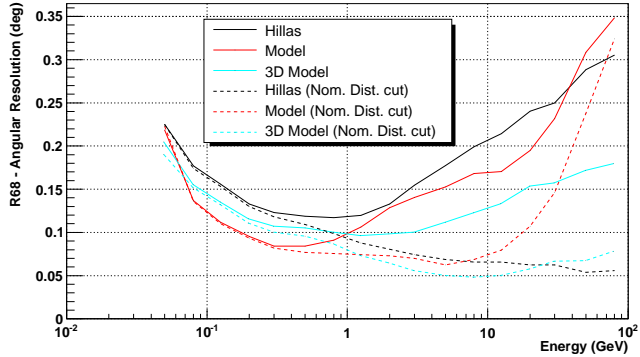,height=5cm}
\caption{Angular resolution of the three analysis methods as function of
energy, with and without nominal distance cut.}
\label{fig:Denaurois-figResolution}
\end{center}
\end{figure}

The results of the angular resolution comparison is shown in figure \ref{fig:Denaurois-figResolution},
with and without nominal distance cut. For all analyses, applying a nominal distance cut
rejects the high energy events falling far away from the telescope (which have almost 
parallel images in the telescopes) thus improving the angular resolution at the expense
of a much smaller effective area (by a factor of typically 5 at 10 TeV).
The Model analysis performs significantly better than other analyses at low energy
whereas the 3D Model takes over at higher energies.

\begin{figure}[htb]
\begin{center}
\epsfig{file=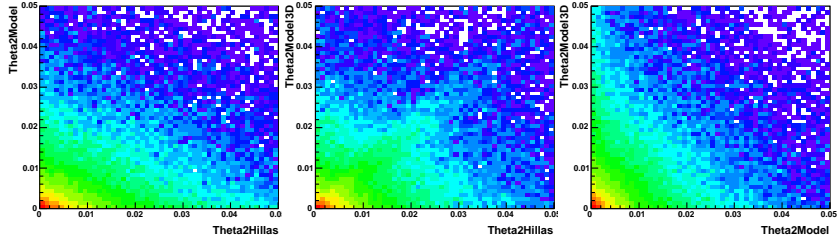,width=11cm}
\caption{Correlation of the reconstructed squared angular distance to true shower axis
($deg^2$) between respectively the Hillas and Model analyses (left), the Hillas and 3D Model
analyses (middle) and the Model and 3D Model analyses (right).}
\label{fig:Denaurois-figAngularCorrelation}
\end{center}
\end{figure}

More interestingly, figure \ref{fig:Denaurois-figAngularCorrelation} shows the correlation
of the reconstructed squared angular distance to shower true direction ($\theta^2$) 
between the three analyses. The values of $\theta^2$ are not very much correlated 
(correlation factors between $0.3$ and $0.5$) between the analyses. This has been identified
to be due to different reconstructions patterns on the ground:  The Hillas analysis best 
reconstructs the events that are well within the array, whereas the Model analysis performs
better with events that are not too close to one telescope. The 3D Model does its best
with high telescope-multiplicity events, which concentrates at the center of the array.

%%%%%%%%%%%%%%%%%%%%%%%%%%%%%%%%%%%%%%%%%%%%%%%%%%%%%%%%%%%%%%%%%%%%%%%%%

\subsection{Off-axis observations}

\begin{figure}[htb]
\begin{minipage}[b]{0.45\linewidth}
\begin{center}
\epsfig{file=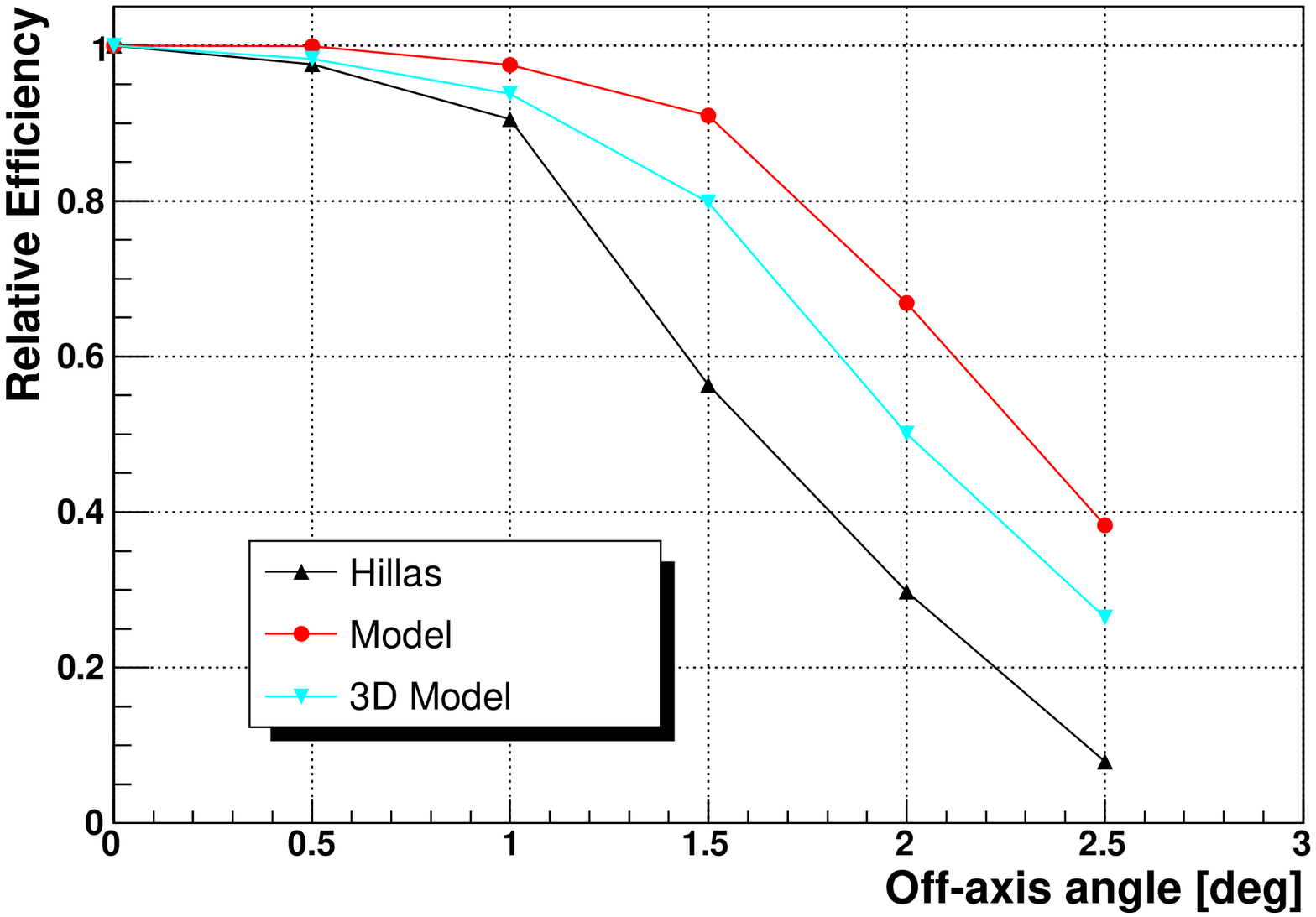,width=5.5cm} 
\caption{Relative efficiency to $\gamma$-rays of the 3 analyses as a function of
distance to camera centre.}
\label{fig:Denaurois-figOffAxisEfficiency}
\end{center}
\end{minipage}
\hspace{0.4cm}
\begin{minipage}[b]{0.45\linewidth}
\begin{center}
\epsfig{file=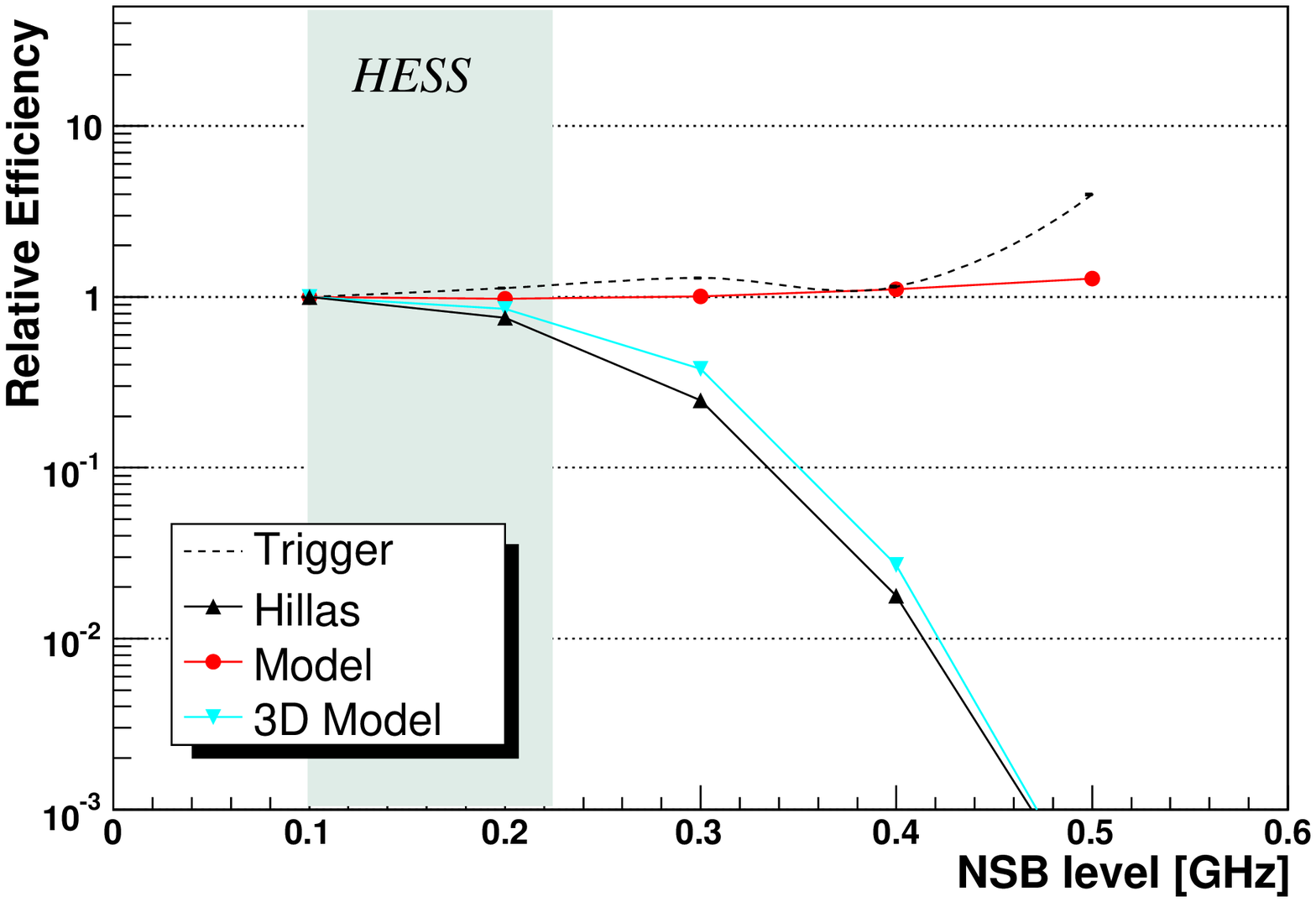,width=5.5cm} 
\caption{Relative efficiency to $\gamma$-rays of the 3 analyses as a function of
night sky background level.}
\label{fig:Denaurois-figNSBEfficiency}
\end{center}
\end{minipage}
\end{figure}

Figure \ref{fig:Denaurois-figOffAxisEfficiency} shows the relative evolution of the
efficiency for $\gamma$-rays as function of the OFF-axis angle (distance of the shower
axis to the centre of the camera) for the three analyses. As expected, the efficiency
of the Hillas analysis starts to fall off before the others: the efficiency of the 
Mean Scaled Width/Length parameters degrades quickly due to truncated showers. Neither 
the Model nor the 3D Model analysis relies on the actual --- truncated --- images but rather 
extrapolate the available information and have therefore a flatter efficiency.

\subsection{Sensitivity to Night Sky Background}

The variation of the efficiency  to $\gamma$-rays with respect
to the Night Sky Background level (NSB) is shown for the three analyses considered
in figure \ref{fig:Denaurois-figNSBEfficiency}. The efficiency of the Hillas 
and the 3D Model analyses both start to drop quickly above 200~MHz, whereas
the efficiency of the Model analysis is much flatter, due to its complete
treatment of the NSB level in the {\it goodness-of-fit} parameter. 

\begin{figure}[htb]
\begin{center}
\epsfig{file=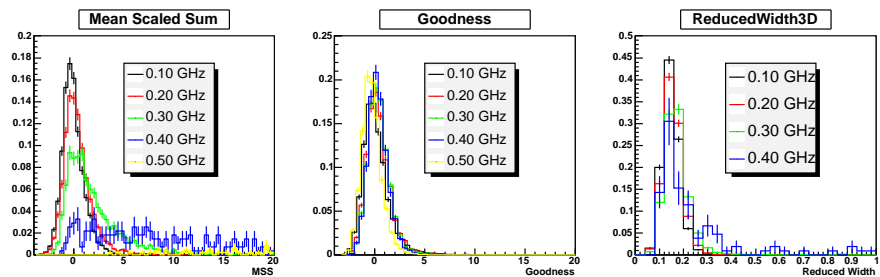,width=12cm}
\caption{Evolution of the three discriminating parameter distributions as a
function of the Night Sky Background level.}
\label{fig:Denaurois-figParametersNSB}
\end{center}
\end{figure}

Figure \ref{fig:Denaurois-figParametersNSB} shows the evolution of the three discriminating parameter 
({\it Mean Scaled Sum}, {\it Goodness-of-fit} and {\it Reduced 3D Width}) distributions as a
function of the Night Sky Background level.
As for the OFF-axis  efficiency, the degradation of the Mean Scaled Width/Length parameters is responsible
for the efficiency drop of the Hillas analysis, whereas the 3D Model seems to suffer
mainly from convergence problems which might be solved in the future. As expected, the Model {\it Goodness-of-fit}
distribution remains stable as the NSB level changes.
It should be noted however that in the operational region of the current telescopes (100 - 200~MHz for HESS),
the efficiency of all three analyses remains almost flat.

%\section*{Acknowledgments}
%This is where acknowledgments for funding institutions etc. should be located.
%Note that there are no section numbers for Acknowledgments, Appendices
%or References.
%Part of the templates and files used draw one's inspiration from
%eConf, FPCP 2003 and Moriond proceedings. We acknowledge
%people who produce them.

\section{Conclusion}

We have presented three completely different analysis methods for Atmospheric 
Cerenkov Telescopes. These three methods show similar efficiencies, although
they are sensitive to different properties of the shower. The
intrinsic capabilities of each analysis (and it particular the hadronic
rejection capabilities) can be combined together to improve the sensitivity of
the analysis. Since these three analyses perform differently in different 
energy and impact parameter domain, more detailed studies should also allow to 
use the select on an event-per-event basis the optimal response and therefore 
improve the quality (angular resolution,...) of the analysis.

\section*{Acknowledgments}

The author would like to thank the members of the H.E.S.S. collaboration
for the fruitful discussions about the different analysis techniques
that are in use within the collaboration. A special thank goes
to Marianne Lemoine-Goumard for her tremendous work in the development
of the 3D Model Analysis.

%%%%%%%%%%%%%%%%%%%%%%%%%%%%%%%%%%%%%
% Label to flag the last page of your contribution
% Replace Yourname by your name starting with a capital letter
%
\label{DenauroisEnd}
 

\begin{thebibliography}{99}

%%
%%  bibliographic items can be constructed using the LaTeX format in SPIRES:
%%    see    http://www.slac.stanford.edu/spires/hep/latex.html
%%  SPIRES will also supply the CITATION line information; please include it.
%%

\bibitem{Denaurois-Hillas-1985} A. Hillas, ``Cerenkov light images of EAS produced by primary gamma'',
{\it Proc. 19nd I.C.R.C. (La Jolla), Vol 3, 445 (1985) }

\bibitem{Denaurois-Hofmann-1999} W. Hofmann {\it et al}, ``Comparison of techniques to reconstruct VHE 
gamma-ray showers from multiple stereoscopic Cherenkov images'', 
{\it Astropart. Phys. {\bf 12}, 135 (1999) }

\bibitem{Denaurois-Daum-1997} A. Daum {\it et al}, ``First results on the performance of the HEGRA IACT array'',
{\it Astropart. Phys. {\bf 8}, 1 (1997) }

\bibitem{Denaurois-DeNaurois-2003} M. de Naurois {\it et al}, ``Application of an Analysis Method Based on a 
Semi-Analytical Shower Model to the First H.E.S.S. Telescope'',
{\it Proc. 28nd I.C.R.C. (Tsukuba), Vol 5, 2907 (2003)}

%\bibitem{Denaurois-Lemoine-2004} M. Lemoine-Goumard {\it et al}, ``Selection and 3D-reconstruction of 
%gamma-ray-induced air showers with H.E.S.S.'',
%{\it Proc. 2nd Gamma-ray International Symposium (Heidelberg), 697 (2004)}

\bibitem{Denaurois-Lemoine-ch2005} M. Lemoine-Goumard {\it et al}, ``3D-reconstruction of gamma-ray showers with a stereoscopic system'',
{\it these proceedings pp.~\pageref{Lemoine-GoumardStart}-\pageref{Lemoine-GoumardEnd}}

\bibitem{Denaurois-Piron-2001} F. Piron {\it et al}, ``Temporal and spectral gamma-ray properties of Mkn~421 above 
250~GeV from CAT observations between 1996 and 2000'',
{\it A\&A {\bf 374}, 895 (2001)}

\bibitem{Denaurois-Reynolds-1993} P.~T. Reynolds {\it et al}, ``Survey of candidate gamma-ray sources at TeV energies 
using a high-resolution Cerenkov imaging system - 1988-1991'',
{\it ApJ {\bf 404}, 206 (1993)}

\bibitem{Denaurois-Mohanty-1998} G. Mohanty {\it et al}, ``Measurement of TeV gamma-ray spectra with the Cherenkov imaging technique'',
{\it Astropart. Phys. {\bf 9}, 15 (1998) }

\end{thebibliography}
\end{document}